\documentclass[a4paper]{article}
\usepackage{Odyssey2020}
\usepackage{epsfig,amssymb,amsmath,graphicx,amssymb,xcolor,color,dsfont, url}
\usepackage[hidelinks]{hyperref}
\usepackage{booktabs}
\usepackage{multirow}
\usepackage{url}
\usepackage{bbding}
\usepackage{pifont}

\newcommand{\x}{\boldsymbol{x}}
\newcommand{\Eta}{\boldsymbol{\eta}}
\newcommand{\y}{\boldsymbol{y}}
\newcommand{\transpose}{^{^{\intercal}}}

\usepackage{booktabs}
\usepackage{multirow}
\usepackage{tabularx}
\usepackage{pstricks}
\newcolumntype{Y}{>{\centering\arraybackslash}X}

\title{LEAP System for SRE19 CTS Challenge - Improvements and Error Analysis}

\name{{Shreyas Ramoji, Prashant Krishnan, Bhargavram Mysore, Prachi Singh, Sriram Ganapathy}}

\address{LEAP Lab, Electrical Engineering, Indian Institute of Science, Bangalore.\\
{\small \tt shreyasr@iisc.ac.in }}
\begin{document}
\maketitle

\begin{abstract}
The NIST Speaker Recognition Evaluation - Conversational  Telephone  Speech  (CTS)  challenge 2019  was  an open  evaluation  for  the  task  of  speaker  verification  in challenging conditions.  In this paper, we provide a detailed  account  of  the  LEAP  SRE  system  submitted  to the CTS challenge focusing on the novel components in the back-end system modeling.  All the systems used the time-delay neural network (TDNN) based x-vector embeddings. The x-vector system in our SRE19 submission used a large pool of training speakers (about 14k speakers).   Following  the  x-vector  extraction,  we  explored  a neural network approach to backend score computation that was optimized for a speaker verification cost.  The system combination of generative and neural PLDA models resulted in significant improvements for the SRE evaluation  dataset.   We  also  found  additional  gains  for  the SRE systems based on score normalization and calibration.  Subsequent to the evaluations, we have performed a detailed analysis of the submitted systems.  The analysis  revealed  the  incremental  gains  obtained  for  different training dataset combinations as well as the modeling methods.

\end{abstract}

\section{Introduction}
\label{sec:intro}
The recent years have seen increasing demand for authentication and verification systems using speech. In defense applications, speaker detection is an important aspect in surveillance of telephone recordings while in commercial applications like banking, voice-operated smart speakers and mobile phones, the use of speech based authentication is becoming ubiquitous. The acceptable performance of the system relies on relatively clean recordings and with matched languages used in training and testing the systems. The performance is substantially degraded in noisy and multi-lingual environments making the downstream applications vulnerable. Over the past two decades, the NIST  speaker recognition evaluation (SRE) challenges provide a suitable benchmark for comparing and standardizing speaker recognition systems. The NIST Speaker Recognition Evaluation 2019 \cite{Sadjadi19plan} is the latest among an ongoing series of challenges, and it consisted of two tracks - the first was a leaderboard style evaluation on speaker detection from  Conversational Telephone Speech (CTS), and the second one was a multimedia speaker recognition. This paper reports the efforts of the LEAP system submission to the SRE19 CTS challenge which advance our previous efforts on the SRE18 challenge \cite{ramoji2019leap}.

The conventional approach deployed for speaker recognition consisted of the Gaussian mixture modeling (GMM) of speech training data followed by an adaptation using maximum-aposteriori (MAP) principles  \cite{reynolds2000speaker}. The adapted model is compared with the background GMM model using the log-likelihood ratio score.  This approach was advanced by the development of i-vectors as fixed dimensional front-end features for speaker recognition tasks  \cite{kenny2007joint,dehak2011front}.  The i-vectors capture long term information of the speech signal such as speaker and language. In the recent past, the i-vectors derived from deep neural network (DNN) based posterior features were also explored for SID \cite{lei2014novel}. The use of bottleneck features for front-end feature extraction derived from a speech recognition acoustic model has also shown good improvements for speaker recognition \cite{sadjadi2016ibm}. 

Recently, neural network embedding extractor trained for a supervised speaker discrimination task has shown improvements over the i-vector approach. This uses a time delay neural network (TDNN) with a sequence summary layer followed by feed-forward neural network layers that map to the target layer of training speaker classes. The output of the first feed-forward layer following the sequence summary layer is used as embeddings (x-vectors) for speaker recognition \cite{snyder2018x}.  Following the extraction of x-vectors/i-vectors, different speaker verification systems make use of discriminative/generative models in the back-end for computing the scores. The most popular approaches for scoring include support vector machines (SVMs) \cite{campbell2006support,cumani2013pairwise}, Gaussian back-end model \cite{cumani2014generative,ramoji2019leap} and the probabilistic linear discriminant analysis (PLDA) \cite{kenny2010bayesian,jiang2012plda}. Some efforts on pairwise generative and discriminative modeling are discussed in \cite{cumani2013pairwise,cumani2014large,cumani2014generative}.

In this paper, we describe the LEAP submission to SRE19 challenge. All the submitted systems were based on the x-vector approach derived from extended TDNN (E-TDNN) models. We use three different E-TDNN models which were trained with various subsets of the training data. The x-vector extraction was followed by a back-end modeling.  The vanilla system developed for SRE19 used the conventional approach to back-end modeling for the x-vectors. This consisted of various normalization and dimensionality reduction techniques like the Within Class Covariance Normalization \cite{hatch2006within}, length normalization \cite{garcia2011analysis}, Linear Discriminant Analysis (LDA), Local pairwise LDA \cite{he2018local}.  These normalized and dimensionality reduced x-vectors were modelled with PLDA for computing log-likelihood ratios. 

For the back-end modeling, we explored a full neural network approach where the LDA, WCCN and length-normalization were incorporated as shared layers in a Siamese neural network \cite{ramoji2020pairwise}. This neural network operates on a pair of x-vectors  and outputs a verification score. The NPLDA architecture  was designed to perform the equivalent operations involved in the conventional PLDA back-end. A linear layer is used to replicate centering and LDA transformation, a length-norm layer for unit length normalization, and a quadratic layer which replicates the PLDA score (LLR) formula. However, the advantage of the proposed neural back-end is the ability to train the model using a discriminative cost function. The objective function was constructed to minimize the normalized minimum detection cost function $C_{min}$, similar to the efforts in \cite{Mingote2019}. Finally, following the score generation, we performed calibration and fusion of various systems using the techniques similar those in our previous SRE18 submission \cite{ramoji2019leap}.

The rest of the paper is organized as follows: The SRE19 dataset details, the cost metric and training datasets used in our submission are detailed in Section ~\ref{sec:sre19}. In Section \ref{sec:xvector}, we give an overview of the x-vector feature extraction in three different training configurations using sub-sets of the training data. The back-end approaches using a traditional approach as well as the proposed neural network approach are detailed in Section \ref{sec:backend}. The experiments on the individual systems and calibration and fusion developed for SRE19 are reported in Section~\ref{sec:exp}.  This is followed by Section \ref{sec:results} where we report the analysis and improvements in the post-eval experiments. Finally,  a summary of the paper is provided in Section \ref{sec:summary}.

\section{SRE19 : Datasets and cost metric}\label{sec:sre19}
Given a segment of speech and the target speaker enrolment data, the speaker verification task is to automatically determine whether the target speaker is present in the test segment. A test segment along with the enrolment speech segment(s) constitutes a \emph{trial}. The system is required to process each trial independently and to output a log-likelihood ratio (LLR) score for that trial. The LLR for a given trial including a test segment $u$ is defined as follows:
\begin{align}
    LLR(u) &= \log \left( \frac{P(u|H_0)}{P(u|H_1)} \right)
\end{align}
where $P()$ denotes the probability density function (pdf), and $H_0$ and $H_1$ represent the null (i.e., $u$ is spoken by the enrolled speaker) and alternative (i.e., $u$ is not spoken by the enrolled speaker) hypotheses, respectively.


\begin{table*}[t!]
\begin{center}
\caption{Details of the training and development datasets used in the SRE19 evaluation. We indicate the data partitions used in the three x-vector systems (XV1-XV3) and the back-end models of the seven individual systems submitted (A-G).}
\vspace{0.1cm}
\centering
\resizebox{\textwidth}{!}{
\begin{tabular}{@{}l|ccc|ccc|ccccccc@{}}
\toprule
\multirow{2}{*}{Dataset} & \multirow{2}{*}{\# Speakers} & \multirow{2}{*}{\# Utterances} & \multirow{2}{*}{\#Hours} & \multicolumn{3}{c|}{X-Vector Training} & \multicolumn{7}{c}{Backend Training} \\ \cmidrule(l){5-14} 
                         &                              &                                &                          & XV1            & XV2           & XV3           & A    & B    & C    & D    & E    & F   & G   \\ \midrule
{Voxceleb 1 \& 2}        & 7323                         & 1276888                        & 2781                     & \ding{51}      &               & \ding{51}     &      &      &      &      &      &     &     \\ \midrule
{Mixer 6}                & 591                          & 187197                         & 2068                     &                & \ding{51}     & \ding{51}     &      & \ding{51}     &      &      &      & \ding{51}    &     \\
{SRE 04-06}              & 2238                         & 13346                          & 1114                     &                & \ding{51}     & \ding{51}     &      & \ding{51}      &      & \ding{51}     &      & \ding{51}    &     \\
{SRE 08}              & 1336                         & 9640                          & 684                     &                & \ding{51}     & \ding{51}     & \ding{51}     & \ding{51}      &      & \ding{51}     &      & \ding{51}    &\ding{51}     \\
{SRE 10}              & 446                         & 15561                          &  1272                    &                & \ding{51}     & \ding{51}     &      & \ding{51}      &      & \ding{51}     &      & \ding{51}         \\
{Switchboard Corpus}     & 2594                         & 28181                          & 2457                     &                & \ding{51}     & \ding{51}     &      & \ding{51}      &      &    \ding{51}  &      &\ding{51}     &     \\
{SRE 16 evaluation}      & 201                          & 10496                          & 256                      &                & \ding{51}     & \ding{51}     & \ding{51}     & \ding{51}     &      &      &    \ding{51}  &    \ding{51} &     \\
{SRE18 evaluation}      & 188                          & 13451                          & 258                      &                & \ding{51}     & \ding{51}     &      & \ding{51}     & \ding{51}     &      &\ding{51}      &\ding{51}     & \ding{51}    \\
\bottomrule
{SRE18 dev labelled}    & 25                           & 1741                           & 34                       &                & \ding{51}     & \ding{51}     & \multicolumn{7}{c}{Fusion, Calibration}      \\ 
{SRE18 dev unlabelled}  & -                            & 2332                           & 72                       &                &               &               &    \multicolumn{7}{c}{Score Normalization (as-norm) }    \\
\bottomrule
\end{tabular}}
\label{tab:training_dataset}
\end{center}
\end{table*}

The test segment can range from 10 to 60 seconds, and all the trials are gender matched. For a given application, a decision is made by applying a certain application specific threshold to the log-likelihood ratio. 

\subsection{Performance metrics}
The normalized detection cost function (DCF) is defined as
\begin{align}
    C_{Norm}(\beta,\theta) = P_{Miss}(\theta) + \beta P_{FA}(\theta) \label{eqn:cnorm} 
\end{align}
where $P_{Miss}$ and $P_{FA}$ are the probability of miss and false alarms computed by applying detection threshold of $\theta$ to the log-likelihood ratios. The primary cost metric of the NIST SRE18 for the Conversational Telephone Speech (CTS) is given by
\begin{align}
    C_{primary} = \frac{1}{2}\left[C_{Norm}(\beta_1, \log \beta_1) + C_{Norm}(\beta_2, \log \beta_2)\right] \label{eqn:Cprimary}
\end{align}
where $\beta_1 = 99$ and $\beta_2 = 199$.

The minimum detection cost (minDCF or $C_{min}$) is computed by using the detection thresholds that minimize the detection cost.
\begin{align}
    C_{min} = \underset{\theta_1,\theta_2}{\min}\,\,\frac{1}{2}\left[C_{Norm}(\beta_1, \theta_1) + C_{Norm}(\beta_2, \theta_2)\right]
\end{align}

The Equal Error Rate (EER) is the value of $P_{FA}$ or $P_{Miss}$ computed at the threshold where $P_{FA} = P_{Miss}$. We report the results in terms of EER, $C_{Min}$ and $C_{primary}$ for all our systems.

\subsection{Dataset}
The training and development datasets used in our systems is summarized in Table~\ref{tab:training_dataset}. This choice of datasets is based on fixed condition training requirements mentioned in the evaluation plan~\cite{Sadjadi19plan}. The Voxceleb dataset and the previous SRE evaluation datasets along with the Switchboard corpus represent the major components of the training data. We had three variants of x-vector model training.
 While the Voxceleb dataset is primarily used for x-vector training, the other datasets were used for back-end (PLDA) training. The total duration of these datasets is about $11$k hours of speech and it includes about $15$k speakers. The SRE18 development set is held out from rest of the training for model fusion and hyper-parameter selection. The score normalization and calibration used the SRE18 unlabeled dataset. The SRE19 evaluation consisted of $2,688,376$ trials from $14,561$ segments. The mean duration of the recordings was around $60$ seconds.  More details about the SRE19 evaluation data can be found in the evaluation plan~\cite{Sadjadi19plan}.

\section{Front-end modeling}\label{sec:xvector}

We trained three x-vector models with different subsets of the training data using the extended time-delay neural network architecture described in \cite{snyder2019speaker}. For x-vector extraction, an extended TDNN with $12$ hidden layers and rectified linear unit (RELU) non-linearities is trained to discriminate among the speakers in the training set. The first $10$ hidden layers operate at frame-level, while the last $2$ operate at segment-level.   There is a $1500$-dimensional statistics pooling layer with between the frame-level and segment-level layers that accumulates all frame-level  outputs  from  the  $10$th layer  and  computes  the  mean  and  standard  deviation  overall frames for an input segment.  After training, embeddings are extracted from the $512$-dimensional affine component of the $11$th layer (i.e., the first segment-level layer).  More details regarding the DNN architecture (e.g., the number of hidden units per layer) and the training process can be found in \cite{snyder2019speaker}.

\subsection{VoxCeleb x-vector system (XV1)}\label{sec:voxxvector}

\subsubsection{Training Datasets}
The x-vector extractor is trained entirely using speech data extracted from combined VoxCeleb 1 and 2 corpora~\cite{nagrani2017voxceleb}. These datasets contain speech extracted from celebrity interview videos available on YouTube, spanning a wide range of different ethnicity, accents, professions, and ages. For training the x-vector extractor, we use about $1.2$M segments from $7323$ speakers selected from VoxCeleb 1 (dev and test), and VoxCeleb 2 (dev). 

\subsubsection{Feature Configuration and Model Description}
The XV1 x-vector extractor was trained using $23$ dimensional Mel-Frequency Cepstral Coefficients (MFCCs) from $25$ ms frames every $10$ ms using a $23$-channel mel-scale filter-bank spanning the frequency range $20$ Hz - $3700$ Hz. In order to increase the diversity of the acoustic conditions in the training set, a 5-fold augmentation strategy is used that adds four corrupted copies of the original recordings to the training list. The recordings are corrupted by either digitally adding noise (i.e., babble, general noise, music) or convolving with simulated and measured room impulse responses (RIR). The noise and RIR samples are freely available\footnote{ http://www.openslr.org}. Augmenting the original data with the noisy versions gives 
$6.3$M training segments for the combined VoxCeleb dataset.

\subsection{SRE x-vector system (XV2)}\label{sec:srexvector}

\subsubsection{Training Datasets}
The x-vector extractor is trained using speech data extracted from SwitchBoard corpus, Mixer 6, SRE04-10, SRE16 evaluation set and SRE18 development and evaluation sets. We used with $0.5$M recordings from $6217$ speakers. The datasets were augmented with the $5$-fold augmentation strategy similar to the previous model. The recordings are corrupted by either digitally adding noise (i.e., babble, general noise, music) or convolving with simulated and measured room impulse responses (RIR). 

\subsubsection{Feature Configuration and Model Description}
This x-vector model used $30$ dimensional MFCC features using a $30$-channel mel-scale filterbank spanning the frequency range $200$ Hz - $3500$ Hz. 
All other hyper-parameters were the same as the XV1 x-vector system.
The E-TDNN x-vector system was trained using speakers that had more than $8$ utterances per speaker.  

\subsection{Full X-Vector System (XV3)}\label{sec:megaxvector}

\subsubsection{Training Datasets}
By combining the Voxceleb 1\&2 dataset with Switchboard, Mixer 6, SRE04-10, SRE16 evaluation set and SRE18 development and evaluation sets, we obtained with $2.2$M  recordings from $13539$ speakers. The datasets were augmented with the 5-fold augmentation strategy similar to the previous models. In order to reduce the weighting given to the VoxCeleb speakers (out-of-domain compared to conversational telephone speech (CTS)), we also subsampled the VoxCeleb augmented portion to include only $1.2$M utterances.

\subsubsection{Feature Configuration and Model Description}
This x-vector model uses $30$ dimensional MFCCs using a $30$-channel mel-scale filterbank spanning the frequency range $20$ Hz - $3700$ Hz. 
All other hyperparameters were kept the same as the first x-vector system.

\begin{table*}[t!]
\caption{Impact of adding in-domain and out-of-domain data in the training of backend models. The front-end x-vectors for these models come from the XV1 system.}
\vspace{0.25cm}
\centering
\label{table:nplda_analysis}
\begin{tabular}{@{}l|l|lc|cc|cc@{}}
\toprule
\multirow{2}{*}{Model} & \multirow{2}{*}{Train Datasets} & \multicolumn{2}{c|}{SRE18 Dev} & \multicolumn{2}{c|}{SRE18 Eval} & \multicolumn{2}{c}{SRE19 Eval} \\ \cmidrule(l){3-8} 
 &  & EER (\%) & $C_{Min}$ & EER (\%) & $C_{Min}$ & EER (\%) & $C_{Min}$ \\ \midrule
GPLDA & SRE 04-10, SWBD, MX6 & 11.4 & 0.65 & 13.2 & 0.70 & 13.3 & 0.69 \\
GPLDA & + SRE16 & 10.0 & 0.58 & 11.4 & 0.65 & 12.2 & 0.66 \\
GPLDA & + SRE18 Eval & \textbf{8.73} & 0.56 & 6.93 & 0.51 & 9.63 & 0.58 \\ \midrule
NPLDA & SRE 04-10, SWBD, MX6 & 10.8 & 0.60 & 10.1 & 0.64 & 10.7 & 0.64 \\
NPLDA & + SRE16 & 10.9 & 0.53 & 9.61 & 0.61 & 10.4 & 0.63 \\
NPLDA & + SRE18 Eval & 9.53 & \textbf{ 0.49} & \textbf{6.73} & 
\textbf{0.48} & \textbf{8.64} & \textbf{0.55 } \\ \bottomrule
\end{tabular}
\end{table*}

\section{Back-end modeling}\label{sec:backend}

\subsection{Generative PLDA (GPLDA)}\label{sec:kaldiplda}
The primary baseline we use to benchmark our systems is the Probabilistic Linear Discriminant Analysis (PLDA) \cite{sizov2014unifying} back-end implementation in the Kaldi toolkit \cite{povey2011kaldi}. This PLDA model is based  on the two-covariance modeling approach.  In order to train model, the x-vectors are centered, dimensionality reduced using Linear Discriminant Analysis (LDA), followed by unit length normalization~\cite{garcia2011analysis}. These processed x-vectors are then used to train the PLDA model.

During the training, the GPLDA implementation computes a linear transform to center and simultaneously diagonalize the within and between class covariance of the training embeddings. These pre-processing steps are summarized as follows:
$$ \underset{\text{x-vector}}{\x_r} \xrightarrow{\substack{\text{Centering,}\\ \text{LDA}}}  \y_r \xrightarrow{\substack{\text{Unit Length}\\ \text{Normalization}}} \hat{\y}_r \xrightarrow{\substack{\text{Diagonalizing}\\ \text{Transform}}} \underset{\substack{\text{pre-processed}\\ \text{embedding}}}{\Eta_r}$$

The PLDA model on the processed x-vector for a given recording  is,
\begin{equation}
\Eta _r = \Phi \boldsymbol{\omega} + \boldsymbol{\epsilon}_r 
\end{equation}
where $\Eta _r$ is the x-vector for the given recording, $\boldsymbol{\omega}$ is the latent speaker factor with a prior of $\mathcal{N}(0,I)$, $\Phi$ characterizes the speaker sub-space matrix. The across class covariance matrix (which captures across speaker variability) is denoted by $\Sigma_{ac} = \Phi \Phi \transpose$. $\boldsymbol{\epsilon}_r$ is the residual term with distribution $\mathcal{N}(0,\Sigma_{wc})$ which is intended to capture session  variability such as language, channel, noise, etc.

We denote the pre-processed embeddings of the enrolment and test segments as $\Eta_e$ and $\Eta_t$ respectively. The PLDA log-likelihood ratio is computed as
\begin{equation}\label{eq:plda_scoring}
s(\Eta_e, \Eta_t) = \Eta_e\transpose Q \Eta_e + \Eta_t\transpose Q \Eta_t + \Eta_e\transpose P \Eta_t + c
\end{equation}
where, 
\begin{eqnarray}
Q = \Sigma _{tot} ^{-1} -  (\Sigma _{tot} - \Sigma _{ac} \Sigma _{tot}^{-1} \Sigma _{ac})^{-1} \\
P =  \Sigma _{tot} ^{-1} \Sigma _{ac} (\Sigma _{tot} - \Sigma _{ac} \Sigma _{tot}^{-1} \Sigma _{ac})^{-1}
\end{eqnarray}
with $\Sigma _{tot} = \Sigma_{ac}  + \Sigma_{wc}$. Here, $c$ is a constant term independent of the trial arising from the parameters of the latent variable distributions.

\subsection{Neural PLDA}\label{Neural PLDA}
\begin{figure}[t]
    \includegraphics[width=\linewidth, trim={2.2cm 2.1cm 1.5cm 1.5cm},clip]{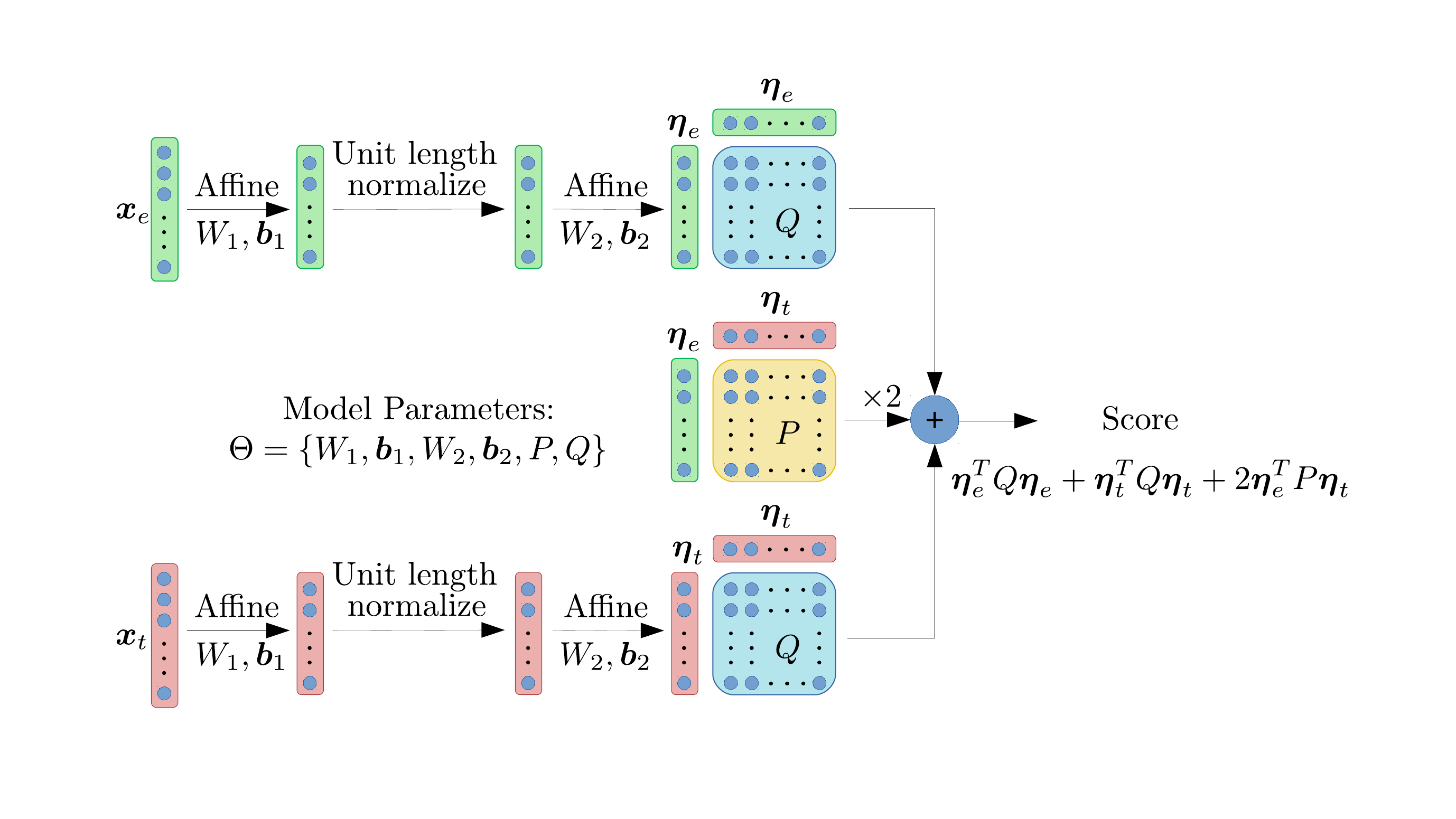}
    \caption{NPLDA model architecture: The two inputs $\x_1$ and $\x_2$ are the enrollment and test x-vectors which constitute a \textit{trial}.}
    \label{fig:PldaNet}
    \vspace{-2ex}
\end{figure}

In the proposed pairwise discriminative PLDA model (neural PLDA), we pose the pre-processing steps and the log-likelihood ratio computation steps in the generative modeling as a   function learnable in a neural network framework (Fig.~\ref{fig:PldaNet}).  Specifically,  we implement the pre-processing steps of centering and LDA as an affine layer. The unit-length normalization is implemented as a non-linear activation and PLDA centering and diagonalizing transform is implemented as another affine layer. Finally, the PLDA log-likelihood ratio given in Eq.~\ref{eq:plda_scoring} is implemented as a quadratic layer as shown in Fig.~\ref{fig:PldaNet}. Thus, the neural PLDA (NPLDA) implements the pre-processing of the x-vectors and the PLDA scoring as a neural back-end. The model parameters of the NPLDA are initialized with the baseline system and these parameters are learnt in a backpropagation setting.

\subsubsection{Loss Function}
The probability of miss and false alarms in Eq.~\ref{eqn:cnorm} computed by applying a detection threshold $\theta$ are, 
\begin{align}
    P_{Miss}(\theta) &= \frac{\sum_{i=1}^{N} t_i \mathds{1}(s_i<\theta)}{\sum_{i=1}^{N} t_i}\\
    P_{FA}(\theta) &= \frac{\sum_{i=1}^{N} (1-t_i) \mathds{1}(s_i \geq \theta)}{\sum_{i=1}^{N} (1-t_i)}
\end{align}
Here, $s_i$ is the score output by the model, $t_i$ is the ground truth variable for trial $i$. That is, $t_i = 0$ if trial $i$ is a target trial, and $t_i = 1$ if it is a non-target trial. $\mathds{1}$ is the indicator function. The normalized detection cost function (Eq.~\ref{eqn:cnorm}) is not a smooth function of the parameters due to the step discontinuity induced by the indicator function $\mathds{1}$. We propose a differentiable approximation of the normalized detection cost by approximating the indicator function with a warped sigmoid function similar to the efforts in \cite{Mingote2019} applied for text dependent end-to-end speaker verification. 
\begin{align}
    P_{Miss}^{\text{(soft)}}(\theta) &= \frac{\sum_{i=1}^{N} t_i  \left[1-\sigma(\alpha(s_i-\theta))\right]}{\sum_{i=1}^{N} t_i} \label{eqn:pmisssoft}\\
    P_{FA}^{\text{(soft)}}(\theta) &= \frac{\sum_{i=1}^{N} (1-t_i) \sigma(\alpha(s_i - \theta))}{\sum_{i=1}^{N} (1-t_i)} \label{eqn:pfasoft}
\end{align}
By choosing a large enough value for warping factor $\alpha$, the approximation can be made arbitrarily close to the actual detection cost function for a wide range of thresholds.

We approximate $P_{Miss}$ and $P_{FA}$ terms in the primary cost metric (Eqn. \ref{eqn:Cprimary}) of the NIST SRE18 (CTS) with their soft counterparts to obtain a differentiable loss function
\begin{align}
     \mathcal{L}_{Primary} = \frac{1}{2}\left[C_{Norm}^{\text{(soft)}}(\beta_1, \theta_1) + C_{Norm}^{\text{(soft)}}(\beta_2, \theta_2)\right]
     \label{eqn:softloss}
\end{align}

We train the pairwise NPLDA model with this differentiable cost function computed over gender matched trials. The proposed loss function in Eq.(~\ref{eqn:softloss})  is novel compared to previous attempts at discriminative modeling for speaker recognition using triplet loss or binary cross entropy loss.

\subsubsection{Sampling of Trials and NPLDA Training}
The procedure to sample trials is similar to what we used for the pairwise Gaussian Back-end model in our previous work \cite{ramoji2019leap}. We randomly sample pairs of gender matched x-vectors from each dataset belonging to target and non-target trials. Along with these manually sampled trials, we also include the SRE08,SRE10 and SRE16 evaluation trials from conversational telephone speech condition. This generates a total of $5$M trials for the NPLDA training. 

Unlike the cross entropy loss which is the negative log-likelihood of the labels, the soft DCF requires estimating $P_{Miss}^{\text{(soft)}}$ and $P_{FA}^{\text{(soft)}}$ for each batch. As any imbalance in target to non-target trial ratio in the mini-batches impacts the NPLDA model training,  we choose a large batch size for training the NPLDA  network. The implementation of the NPLDA can be found here\footnote{ \url{https://github.com/iiscleap/NeuralPlda}}.

\begin{table*}[t]
%
\caption{Performance of the individual systems developed for SRE19 evaluation and the fusion system. The individual system results were obtained with the help of the keys for SRE19 provided by NIST after the evaluations. The best individual system is also highlighted. The description of systems A-G can be found in Table \ref{tab:training_dataset}.}
\vspace{0.1cm}
\centering
\resizebox{\textwidth}{!}
{\begin{tabular}{@{}l|c|c|c|cc|cc@{}}
\toprule
\multirow{2}{*}{System} & \multirow{2}{*}{Front-end} & \multirow{2}{*}{Backend} & \multirow{2}{*}{Backend Train Datasets} & \multicolumn{2}{c|}{SRE18 Dev} & \multicolumn{2}{c}{SRE19 Eval} \\ \cmidrule(l){5-8} 
 &  &  &  & EER (\%) & $C_{Min}$ & EER (\%) & $C_{Min}$ \\ \midrule
A & XV1 & GPLDA & SRE16, SRE08 & 10.6 & 0.6 & 11.1 & 0.66 \\
B & XV3 & NPLDA & SWBD, SRE(04-16), MX6, SRE18 Eval & \textbf{5.31} & \textbf{0.32} & \textbf{4.97} & \textbf{0.42} \\
C & XV1 & GPLDA & SRE18 Eval & 7.61 & 0.48 & 7.36 & 0.55 \\
D & XV2 & GPLDA & SWBD, SRE(04-10) & 10.2 & 0.56 & 11.7 & 0.66 \\
E & XV3 & GPLDA & SRE16, SRE18 Eval & 6.07 & 0.38 & 5.81 & 0.45 \\
F & XV3 & GPLDA & SWBD, SRE(04-16), MX6, SRE18 Eval & 7.1 & 0.44 & 7.04 & 0.50 \\
G & XV3 & GPLDA & SRE08, SRE18 Eval & 6.87 & 0.39 & 5.65 & 0.43 \\ \midrule
B+C & - & - & - & - & - & 4.43 & 0.38 \\
B+G & - & - & - & - & - & 4.18 & 0.36 \\ \bottomrule
\end{tabular}}
\label{tab:system_performance}
\end{table*}

\subsection{Comparing Backend Models}
Using the same front-end embedding extractor (XV1 configuration), we compare the two backend approaches based on generative GPLDA model as well as the neural PLDA model. These results are reported in Table~\ref{table:nplda_analysis}. 
  We train GPLDA models with Kaldi using SRE 04-10, switchboard corpus and Mixer6. To this, we add SRE16 and SRE18 data to study the improvements of adding more data. We then initialize the NPLDA model using the GPLDA back-end parameters and retrain the model using the cost function proposed. Table \ref{table:nplda_analysis} summarizes the performance of these systems. In all the cases, the NPLDA yields significant improvements over PLDA. The improvements are more significant for the SRE cost function ($C_{Min}$) as the model is optimized for that metric. 

\section{Systems Submitted}\label{sec:exp}
As mentioned in the previous section, we had three different x-vector extraction models and two different back-end modeling approaches. In addition, several subsets of the training data were optionally used in the backend training. The overview of these systems is given in Table~\ref{tab:training_dataset}.

\begin{itemize}
\item{System A:}
We use the XV1 x-vectors. The GPLDA model is trained using SRE16 eval set and the SRE08 dataset.

\item{System B:}
We use the x-vectors from full x-vector (XV3) model and train a NPLDA model using the Switchboard, Mixer 6 and SRE datasets, including SRE 04-10, 16 and 18. We apply a sigmoid non-linearity at the output and optimize the proposed soft detection loss function. 

\item{System C:}
We use the XV1 x-vectors with a GPLDA model trained with SRE18 evaluation set.

\item{System D:}
The XV2 x-vectors were used along with  the GPLDA model trained using the Switchboard and SRE datasets.

\item{System E:}
We use the x-vectors from full XV3 system and the  GPLDA model is trained using SRE16 and SRE18 evaluation dataset.

\item{System F:}
We use the x-vectors from the XV3 model and the GPLDA model is trained using the Switchboard, Mixer 6 and SRE datasets, including SRE 04-10, 16 and SRE18 evaluation dataset.

\item{System G:}
We use the x-vectors from the XV3 model and the GPLDA model is trained using the  SRE08 dataset, SRE18 evaluation dataset only. 
\end{itemize}

\subsection{Calibration and Fusion}\label{sec:calib} 
A linear score fusion of the different systems is done using the FoCAL toolkit \cite{brummer2007focal},  where the weights and biases are obtained with a logistic regression objective using a held-out set (SRE18 development set). 
In our experiments, we performed fusion of System B (NPLDA) and C, and the fusion of System B and G using the above mentioned approach. The systems for fusion were selected based on the complementary nature of training methods and datasets. The results on SRE19 evaluation set using the fused system scores are listed in Table \ref{tab:system_performance}.

For SRE19 submission, we attempted to calibrate the scores of the final systems using an affine transform which normalizes the within class score variance.  The scores were then mean shifted such that the threshold corresponding to the minimum cost was moved to the target operating point (the operating point for actual cost is given in NIST SRE18 evaluation plan \cite{Sadjadi19plan}). This was performed so as to minimize the difference between $C_{min}$ and $C_{primary}$ on the SRE18 development dataset. This resulted in the $C_{primary}$ for our submission systems to be far from the minimum cost $C_{min}$, and hence we have not reported this in Table~\ref{tab:system_performance}.  In Section \ref{sec:results}, we analyze the issues with this approach for calibration and highlight the steps we have taken to improve the calibration.   

\begin{table*}[t]
\begin{center}
\centering
\caption{Performance of post-eval systems using improved calibration and adaptive score normalization (AS-Norm).}
\vspace{0.25cm}
\label{tab:posteval}
\begin{tabular}{@{}c|c|cc|ccc@{}}
\toprule
\multirow{2}{*}{Model} & \multirow{2}{*}{Train Datasets} & \multicolumn{2}{c|}{SRE18 Dev} & \multicolumn{3}{c}{SRE19 Eval} \\ \cmidrule(l){3-7} 
 &  & EER (\%)& $C_{Min}$ & EER (\%)& $C_{Min}$ & $C_{primary}$ \\ \midrule
GPLDA (XV1) & SRE08, SRE18 Eval & 7.38 & 0.50 & 7.61 & 0.54 & 0.59  \\
 & + AS-NORM & 6.66 & 0.38 & 6.73  & 0.45  & 0.49 \\
GPLDA (XV3) & SRE08, SRE18 Eval & 6.87 & 0.39  & 5.65  & 0.43  & 0.56 \\
 & + ASNORM & 4.86  & 0.31  & 4.83  & 0.37  & 0.42 \\
NPLDA  (XV3) & SWBD, SRE(04-16), SRE18 Eval & 4.88 & 0.28 & 4.56 & 0.39 & 0.47 \\
 & + AS-NORM  & \textbf{4.73}  &\textbf{0.27}  & \textbf{4.51}  & \textbf{0.36 }  & \textbf{0.39}  \\
 \bottomrule
 Fusion B+G & + AS-NORM & - & - & 4.22 & 0.34 & 0.39 \\
 \bottomrule
\end{tabular}
\end{center}
\vspace{-0.2cm}
\end{table*}

\subsection{Summary of Results}
The results obtained for the individual systems is given in Table~\ref{tab:system_performance}. The best individual system was the combination of the XV3 x-vector extractor with the proposed NPLDA model. The full x-vector system (XV3) performs significantly better than the VoxCeleb (XV1) and the SRE (XV2) systems for any choice of back-end. 
The SRE 18 evaluation set is the closest to the SRE18 Dev and SRE19 Evaluation data (Tunisian Language). Comparing systems F and G implies that as we add more out of domain data like the older SRE data, switchboard and Mixer 6 in addition to the in domain (SRE18 Eval) data for PLDA training, the performance starts to degrade. Systems B (NPLDA) is trained with the same data as System F, and it is observed that it models both in-domain and out-of-domain data better than the GPLDA.

\section{Post-eval Experimenents and Analysis}\label{sec:results}

\subsection{Calibration}

\begin{figure}[t]
    \includegraphics[width=\linewidth,  trim={0.2cm 0.1cm 1.1cm 0.9cm},clip]{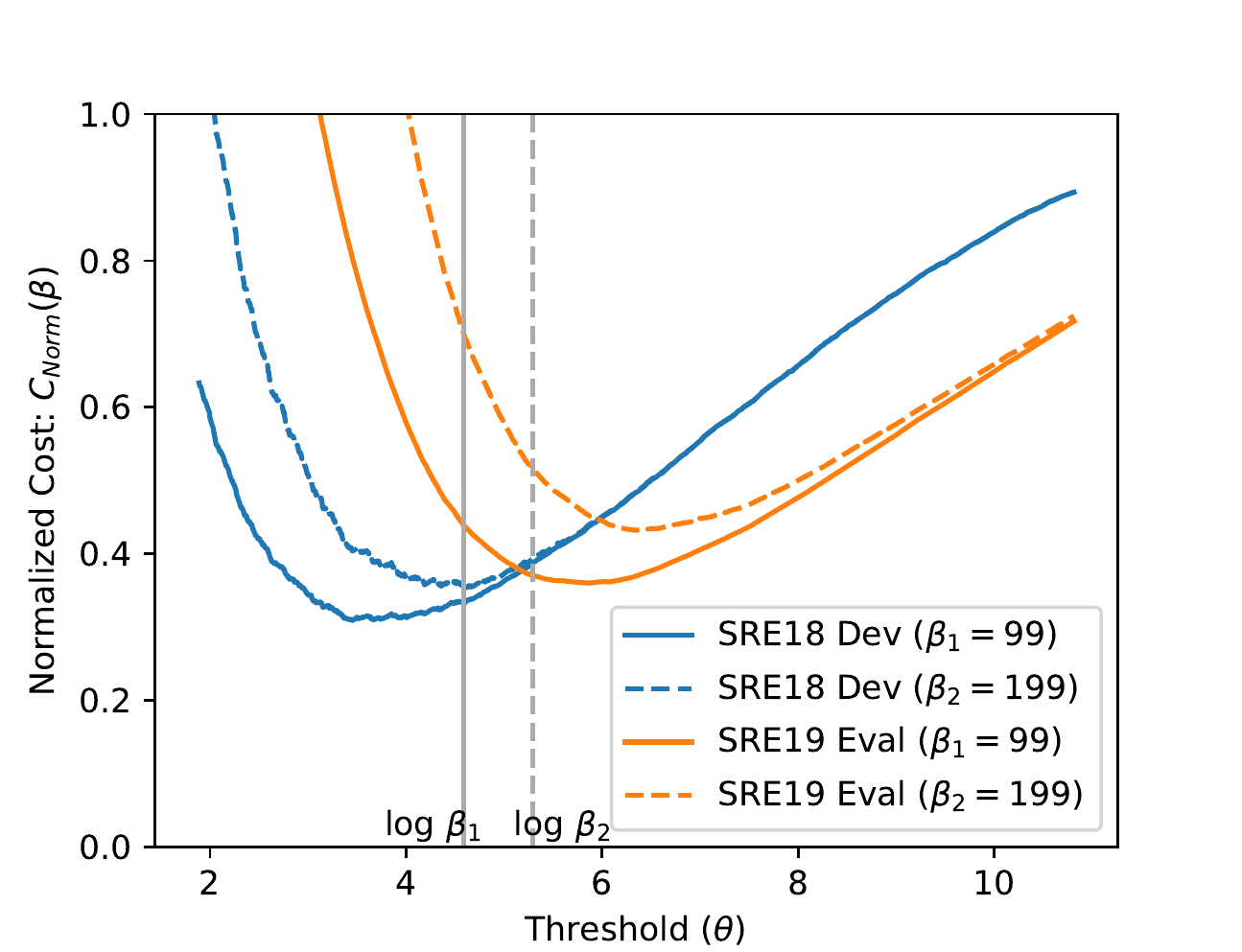}
    \caption{
    Plot illustrating the mismatch in the score distributions in SRE18 dev set and the SRE19 eval set.}
    \label{fig:Dists_costs}
    \vspace{-1ex}
\end{figure}

In our previous work for SRE18 \cite{ramoji2019leap}, we proposed an alternative approach to calibration, where the target and non-target scores are modelled as Gaussian distribution with a shared variance. The calibration procedure in this case involved the shifting of scores so that the threshold corresponding to the minimum cost on the development set is the point where the actual cost is computed on the evaluation trials ($\log{\beta}$ of \cite{Sadjadi19plan}). This was done with the assumption that the score distributions of the development and evaluation trials match closely. Thus, the threshold where $C_{min}$ is achieved in the development set may potentially match with evaluation trials. In the case of SRE18 evaluation~\cite{ramoji2019leap}, the development and evaluation score distributions were more or less the same, and the threshold that minimized the detection cost were very close. However, in SRE19, there was no exclusive matched development dataset provided. Hence, aforementioned calibration method using the SRE18 development dataset apploed on the SRE19 evaluation trials (as done for our submitted systems) turned out to be ineffective. Given the keys for SRE19 evaluation, we performed a score analysis and this is shown in   Figure \ref{fig:Dists_costs}. As seen here, the computation of $C_{primary}$ using the distribution of SRE18 resulted in a sub-optimal calibration of the scores.  In the post-eval efforts, we have performed score calibration based on the approach described in  \cite{brummer2013bosaris}. As seen in the plot, this matched the primary cost metric $C_{primary}$ (actDCF) closely with the minimum cost. The results for other individual systems based on the updated score calibration are reported in Table \ref{tab:posteval}. 

\subsection{Score Normalization}
We experiment with various cohort based normalization techniques \cite{karam2011towards, Matejka2017} using the SRE18 dev unlabelled set as the cohort. The best improvements were observed with the adaptive symmetric normalization (referred to as AS-Norm Type 1 in \cite{Matejka2017}).  We achieve $24$\% relative improvement for the VoxCeleb x-vector system (XV1) and $21$\% relative improvement for the full x-vector system (XV3) on SRE18 development set. We achieve a comparatively lower but consistent improvements of about $15$\% on an average for the SRE19 evaluation set for all the systems. The improved results are summarized in Table \ref{tab:posteval}.

\section{Summary and Conclusions}\label{sec:summary}
In this paper, we provide an account of our efforts for the NIST SRE19 CTS challenge. We train three x-vector extractors and back-end models on different partitions of the available datasets, and report the performance of the individual as well as the fusion systems.

We explore a novel discriminative back-end model (NPLDA) inspired from deep neural network architectures and the generative PLDA model. For this model, we optimize a differentiable loss function constructed to approximate the detection cost function. Using a single elegant architecture targeted to optimize the speaker verification loss, the NPLDA uses a pair of x-vectors to directly generate the score. We provide analysis to show that NPLDA significantly boosts the performance of the system over the GPLDA for various datasets.

We discuss the errors that can be caused by calibration with a mismatched development set, and report the gains that can be achieved by using a cohort based adaptive score normalization technique for various systems.

\ninept
\bibliographystyle{IEEEbib}
\bibliography{Odyssey2020_SRE}

%

\end{document}